\documentclass[preprint]{aastex}
\usepackage{fullpage}
\usepackage{color}

\usepackage{natbib}

\slugcomment{Not to appear in Nonlearned J., 45.}
\shorttitle{Monte Carlo Simulation of COMs formation in cold cores}
\shortauthors{Chang \& Herbst}

\begin{document}
\title{Unified Microscopic-Macroscopic Monte Carlo Simulations of Complex Organic Molecule Chemistry in Cold Cores}

\author{Qiang Chang\altaffilmark{1,2,3}, and Eric Herbst\altaffilmark{4}}
\altaffiltext{1}{Xinjiang Astronomical Observatory, Chinese Academy of Sciences, 150 Science 1-Street, Urumqi 830011, PR China}
\altaffiltext{2}{Key Laboratory of Radio Astronomy, Chinese Academy of Sciences, 2 West Beijing Road, Nanjing 210008, PR China}
\altaffiltext{3}{ Department of Chemistry, University of Virginia, Charlottesville, VA 22904 USA}
\altaffiltext{4}{ Departments of Chemistry, Astronomy, and Physics,
University of Virginia, Charlottesville, VA 22904 USA}

\begin{abstract}
The recent discovery of methyl formate and dimethyl ether in the gas phase of cold cores with temperatures as cold as 10 K challenges our previous
astrochemical models concerning the formation of complex organic molecules. 
The strong correlation between the abundances and distributions  of methyl formate and dimethyl ether
further shows that current astrochemical models may be missing important chemical processes in cold astronomical sources.  
We investigate a scenario in which complex organic
molecules and the methoxy radical can be formed on dust grains via a so-called ``chain reaction'' mechanism, in a similar manner to CO$_2$.
A unified gas-grain microscopic-macroscopic Monte Carlo approach with both normal and interstitial sites for icy grain mantles is used to perform the chemical simulations. Reactive desorption with varying degrees of efficiency 
is included to enhance the non-thermal desorption of species formed on cold dust grains.  
In addition, varying degrees of efficiency for the  surface formation of methoxy are also included. 
The observed abundances of a variety of organic molecules in cold cores can be reproduced in our models. The strong
correlation between the abundances of methyl formate and dimethyl ether in  cold cores can also be explained.
Non-diffusive chemical reactions on dust grain surfaces may play a key role in the formation of some complex organic molecules.
\end{abstract}

\maketitle

\section{Introduction}
Interstellar complex organic molecules (COMs), which are defined as carbon-containing molecules with 6 or more atoms, have
attracted increasing interest in the past few decades~\citep{Herbst2009}.  Although this definition can refer to the 
unsaturated carbon-chain species detected in cold cores, the acronym is  more commonly used to refer to the more 
terrestrial-like molecules most strongly associated with hot cores and corinos. The interest in COMs derives  
not only from the fact that COMs can 
be good probes of physical conditions of astronomical sources, but also because COMs may be the basis of 
prebiotic molecules, which may be related to the origin of life. Although COMs were first observed in hot cores~\citep{Blake1987},  
they are also detected in a variety of other types of astronomical sources in the early phases of star 
formation~\citep{Bottinelli2004,Arce2008, Oberg2010, Bacmann2012, Cernicharo2012, Oberg2015}. 

Progress in the astrochemical modeling of COM formation has been driven by new observations of these species.  COMs
were first believed to be synthesized in the gas phase from precursors formed on cold dust grains
and desorbed into the gas as the temperature of dust grains increased~\citep{Charnley1992}. New observations towards
hot corinos and the Galactic center challenged the original hypothesis  that COMs are formed in the gas phase~\citep{Ceccarelli2000, Requena2007}.  
Moreover, new experimental and 
computational results showed that the gas phase syntheses used in the original models of COMs are not efficient 
enough to explain the abundances of methyl formate and other species in hot cores and corinos~\citep{Horn2004, Geppert2007}.
A commonly accepted formation mechanism, which is consistent with observations in these sources, 
has been developed~\citep{Garrod2006}.  In this approach, 
zeroth generation COMs such as methanol are formed on the growing ice mantles of cold dust grains by one mobile reactant, typically an atom or a diatomic molecule bound sufficiently loosely to a grain to diffuse rapidly even at 10 K,
and another species. More complex species than methanol may also be formed by this mechanism. 
As the temperature increases, photodissociation products of zeroth generation COMs, which are typically
radicals, are able to diffuse on dust grain surfaces.  The more strongly bound of these species do not desorb initially;  thus, they
can recombine  via radical-radical association reactions to produce the first generation COMs, such as methyl formate (HCOOCH$_3$).  Subsequent photodissociation and radical-radical reactions produce second-generation COMs, etc.

As the temperature continues to rise, the COMs desorb thermally into the gas, where they can be detected.  
Other mechanisms for desorption of COMs include shock processes~\citep{Requena2006, Requena2008}, in regions such as the Galactic center not closely 
associated with hot cores, and non-thermal processes operative at low temperatures, such as photodesorption  
~\citep{Oberg2009}.     
The thermal mechanism can explain in a semi-quantitative sense most of the COM observations in hot cores and corinos~\citep{Garrod2006,Garrod2008}.
Most of the radical-radical recombination reactions used in this model, however, have not been studied in the laboratory, 
and complications have been found.  For example, the methoxy radical (CH$_{3}$O), used in the formation of methyl formate 
via surface recombination with the formyl radical (HCO), has not been found on ice surfaces in the synthesis of methanol from  
CO, leading to the suggestion that it may be converted into the isomeric form CH$_{2}$OH and not be available to react 
with other radicals~\citep{Cernicharo2012}.  Of course, laboratory and interstellar time scales for surface reactions 
can be  quite different; nevertheless, more research is clearly needed to confirm the radical-radical mechanism.

 The recent  detection of  methyl formate and dimethyl ether in cold cores with temperatures 
as low as 10 K raises more serious questions about the general validity of the radical-radical mechanism 
to form COMs~\citep{Oberg2010, Bacmann2012, Cernicharo2012, Vastel2014}   
because radicals can diffuse only slowly if at all on ice mantles at 10 K, and so cannot react efficiently with each other.  
It must be noted, however, that the studied sources are complex and the low resolution of single-dish telescopes might 
result in observations that mix both low temperature and higher temperature portions.
In order to explain the recent detection of COMs in cold cores, ~\citet{Vasyunin2013a} suggested that the COMs  are formed by
precursors such as methanol which are first formed on grain surfaces at 10 K 
and then desorb to the gas phase via non-thermal mechanisms such as reactive desorption.  Even so, they found that it was necessary to add more gas-phase reactions such as radiative association to produce reasonable abundances of methyl formate and dimethyl ether.  A more recent approach to the gas-phase chemistry by \citet{Balucani2015} indicates that methyl formate can be formed in sufficient abundance via the gas-phase reaction
\begin{equation}
{\rm  O  +  CH_{3}OCH_{2} \longrightarrow  HCOOCH_{3} + H.}
\end{equation}
Moreover, they show that the CH$_{3}$OCH$_{2}$ radical is formed from dimethyl ether, 
leading to a correlation between the abundances of dimethyl ether and methyl formate.    
This correlation has been observed by ~\citet{Jaber2014} over a wide range of sources with differing conditions.
 However, a common problem of gas phase routes to form dimethyl ether is that gas phase methoxy abundances must be high enough
in order to produce enough dimethyl ether~\citep{Vasyunin2013a}.  Given the low fractional abundance 
of gas phase methoxy ($4.7\times10^{-12}$) observed in B1-b~\citep{Cernicharo2012}, it is unclear
that dimethyl ether can be formed solely in the gas phase in cold cores. For instance, in the
standard model of  \citet{Balucani2015},  where the reactive desorption efficiency is set to be 0.01 and gas phase COM 
formation reactions included, at the time when the calculated fractional abundances of dimethyl ether and methyl formate
fit observations in B1-b, the fractional abundance of methoxy 
 is more than one order of magnitude larger than the observed value.  
Thus, even in cold cores where thermal desorption of COMs is very inefficient, 
COMs may also be formed appreciably on dust grains and then desorb into the gas phase via non-thermal processes.
Recently ~\citet{Ruaud2015} suggested that the abundances of gas phase COMs in cold and dense regions can be
significantly enhanced by introducing Eley-Rideal and complex induced reaction mechanisms on dust grains. However,
the simulated abundances of methyl formate in their models are still more than two orders of magnitude smaller 
than the observed values in both B1-b and L1689b.

\begin{figure}
\centering
\resizebox{15cm}{15cm}{\includegraphics{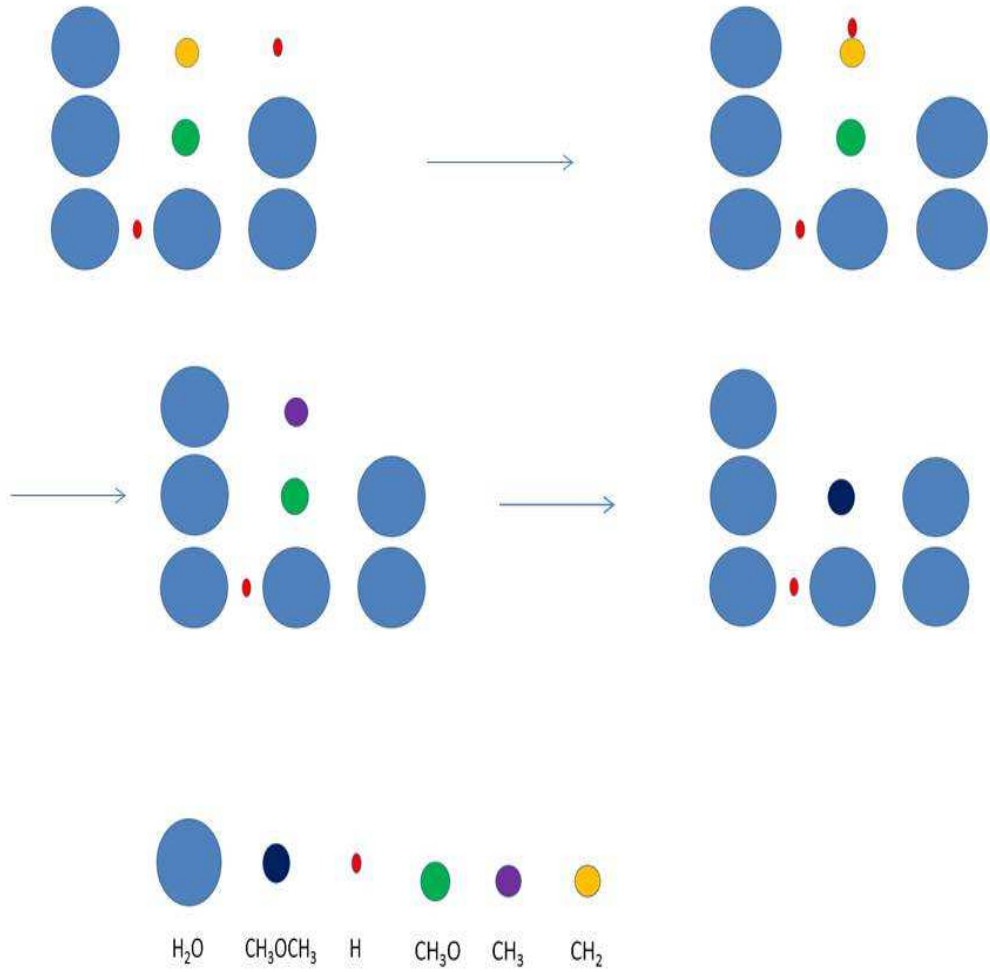}}
\caption{ The chain reaction mechanism leading to the formation of dimethyl ether in a water-rich ice mantle.  See text for a discussion.
}
\label{fig0}
\end{figure}

     In this paper,  we employ a different mechanism to help explain the  detection of COMs in cold sources  
as well as the correlation of the abundances of methyl formate and dimethyl ether. Our explanation is based 
on non-diffusive surface reactions, such as we and others considered for the case of CO$_{2}$, which is one of the major components
of ice mantles~\citep{Oberg2011}.
The most efficient formation mechanism to form CO$_2$ is  the reaction CO + OH $\rightarrow$ CO$_2$ + H, 
although it does possess a significant barrier in the gas. In the laboratory, the surface analog appears 
to proceed via an Eley-Rideal mechanism, in which a gaseous CO molecule lands on and reacts with a surface OH radical, 
formed via photodissociation of water \citep{Yates2014}. 
The efficiency of this reaction in the interstellar medium has not been determined.
Another possible process to enable the CO + OH reaction is the so-called ``chain reaction'' mechanism, 
in which an O atom on a surface reacts diffusively with an H atom to form the OH radical, 
which can subsequently react with CO without undergoing diffusion if a CO molecule lies right below it~\citep{Chang2007,Chang2012,Garrod2011}.  
Without the initial reaction between O and H, the O atoms cannot react efficiently with CO since the process has a significant barrier.  
 Here, as discussed in~\citet{Chang2014},  we use the chain-reaction mechanism to produce COMs on the top layers of ice mantles. 
Note that chain reactions in our model are allowed to occur only if the radical-radical reactions are already in the network, 
so that it is possible that both this mechanism and a diffusive one can be operative, although the role of diffusion is 
limited for the production of COMs at low temperatures.  
In Fig.~\ref{fig0}, we show how dimethyl ether (CH$_{3}$OCH$_{3}$) is formed via a chain reaction in which  an H atom adds to 
a methylene (CH$_{2}$) radical on the surface to form a methyl (CH$_{3}$) radical, which then can react with a methoxy radical (CH$_{3}$O) 
lying beneath the methyl radical.  Note also that the interstitial H atom shown in the figure (see below for a discussion) does not partake in the process.

Recently, we performed a unified microscopic-macroscopic Monte Carlo (UMMMC) simulation using a full gas-grain reaction
network with physical conditions pertaining to cold cores~\citep{Chang2014}. We found that moderate amounts of 
methyl formate and dimethyl ether could be formed on dust grains at temperatures as low as 10 K. 
The major non-thermal desorption mechanism considered was photodesorption; however, it is 
not sufficiently efficient to drive a significant fraction of COMs into the gas phase, so that most methyl formate and dimethyl ether remain on grain surfaces. In this paper, we report calculated gas phase abundances of COMs in a cold core in a simulation that includes  an efficient  
reactive desorption mechanism. In this mechanism, the exothermicity of a surface reaction is used to drive the reaction product into the gas phase.  
In addition to this non-thermal desorption mechanism, we also include a number of improvements to the gas-phase network.  
The improved model is able to reproduce observed abundances of COMs in cold sources as well as the strong correlation 
in abundances between methyl formate and dimethyl ether.

\section{Simulation Methods and Chemical Models}

We confine ourselves to a brief explanation of our simulation methods
because they have been explained in detail in earlier papers~\citep{Chang2005,Cuppen2005,Chang2012,Chang2014}, 
after which our chemical models are introduced.

\subsection{Simulation Methods} 

In our UMMMC approach, the kinetics that occurs on grains is handled by a microscopic approach, 
in which the position of all species is known,  while the gas-phase kinetics is handled by a 
macroscopic approach~\citep{Chang2014}.  In order to simulate chemical kinetics on a dust grain surface with $N$ binding sites, we put
these $N$ sites on a $2L \times 2L$ square lattice where $L^2=N$. Nodes in the lattice that have both even numbered 
$x$ and $y$ coordinates are termed normal while nodes with both odd numbered $x$ and $y$ sites are termed interstitial. 
Normal sites exist both on the surface of a dust grain and in the bulk of the ice, 
while interstitial sites  exist only in the bulk.  
Species on normal sites are called normal while species occupying interstitial sites are labeled
interstitial. Normal species on the topmost (surface) layer can hop from one normal site to 
another or desorb into the gas phase, while interstitial species can diffuse from one 
interstitial site to another in the bulk. 
The surface hopping rate $b_1$ ($s^{-1}$) from one normal site to its nearest neighbor at temperature $T$ 
is determined by the surface diffusion barrier $E_b$ via the expression $b_1= \nu \exp(-E_b/T)$, 
where $\nu$ is the trial frequency. Similarly, the interstitial diffusion
rate within the ice mantle is given by the expression $b_3= \nu \exp(-E_{b2}/T)$ where $E_{b2}$ 
is the interstitial diffusion barrier while the thermal desorption rate is
given by $b_2= \nu \exp(-E_D/T)$ where $E_D$ is the desorption energy. We set $E_b=0.5E_D$ as in ~\cite{Chang2014}. 
 The formation of major species and COMs on dust grain surfaces is found to be independent of the value of the interstitial diffusion barrier, 
$E_{b2}$~\citep{Chang2014}, while the abundance of radicals which are formed in the ice by photodissociation and bulk diffusion of photodissociation
products is strongly dependent on $E_{b2}$. Because we only allow reactive desorption to occur on the topmost layer, 
the value of $E_{b2}$ should not affect the abundance of gas phase COMs when reactive desorption is included in the simulation.
We choose a value, $E_{b2}=0.7E_D$, which ensures that the interstitial diffusion rate is slower than the surface diffusion rate, while
light species such as atomic H can still diffuse within the ice bulk. The desorption energy $E_D$ for each species on ice is taken from
 \citet{Garrod2006} and \citet{Garrod2008}. 

Hopping, interstitial
diffusion, and thermal desorption are all modeled as Poisson processes, so the time interval between two consecutive events 
is given by $\Delta t = -\ln(X)/b_i$ where $X$ is a random number uniformly distributed within (0, 1) while $b_i$ is the rate of
hopping, interstitial diffusion or thermal desorption. A chemical reaction can occur when one reactive species hops into another 
site with a reactive species, or, in the chain reaction mechanism, a second reaction occurs between a newly formed species on the 
surface and one below it. Interstitial species can also diffuse out of the ice mantle 
and become normal surface species~\citep{Chang2014}. 
Photodissociation processes are modeled
as outcomes of another Poisson process, photon arrival. When photons arrive and strike the ice mantle, we keep track of 
the penetration of photons and decide whether photodissociation reactions can happen or not. The exponential nature
of photon penetration is maintained in our simulation~\citep{Chang2014}. Another possible outcome of photon arrival is photodesorption, 
which only occurs on the topmost 2 layers. 

In order to include the reactive desorption mechanism in the UMMMC simulation, whenever a chemical reaction on the topmost layer 
occurs either diffusively or via a chain reaction, we generate a random number $Y$, which is uniformly distributed within (0, 1), and compare $Y$
with a probability $P$. If $Y<P$, then all products of the reaction will desorb into the gas phase, otherwise
these products will remain on the ice mantle. Products of chemical reactions within the ice mantle still remain in 
the bulk, as in~\citet{Chang2014} . The probability $P$ is the efficiency of reactive desorption, which is
highly uncertain. On the other hand, it was found that $P=0.01$ is large enough to explain the observed
gas-phase methanol abundance while $P\leq0.1$ is the highest value considered by ~\citet{Garrod2007}. We report simulation results from 
models with five different $P$ values, 0, 0.01, 0.03, 0.05 and 0.1. 

\subsection{Chemical Models}
The chemical reaction network used is based on the gas-grain reaction network used in~\citet{Chang2014}. 
Networks of this type can simulate both the gas-phase and grain-surface chemistry very well, and predict abundances 
of many classes of molecules accurately. in cold cores.  Therefore, in this paper, we will concentrate on species 
such as COMs not treated by standard simulations, without discussing most other species in any detail. In particular, we emphasize 
methyl formate, dimethyl ether and acetaldehyde, which were found both in B1-b and L1689B \citep{Bacmann2012, Cernicharo2012}, as well as 
 the most ubiquitous COM, methanol.    Although glycolaldehyde has not been found in cold cores so far, we
still report the abundance of glycolaldehyde for comparison. 
Moreover, H$_2$CO is a precursor of HCOOCH$_3$ in some gas phase synthetic routes~\citep{Horn2004,Vasyunin2013a},  while the
simulated abundance of H$_2$CO is more than two orders of magnitude larger than the observed value in B1-b~\citep{Vasyunin2013a},
so we also report the abundance of H$_2$CO in our models.

The cosmic-ray ionization rate is set at $\zeta = 1.3\times 10^{-17}$ s$^{-1}$.
The physical conditions  pertain to the cold regions of dense clouds 
where methyl formate and dimethyl ether have been found~\citep{Bacmann2012}: 
the temperature is fixed at 10 K while the total proton density ($n_{\rm H}$) is $10^5$ cm$^{-3}$.
The dust-to-gas ratio is fixed at $10^{-12}$, and the visual extinction $A_{\rm V} = 10$. 
The major source of photons under these physical conditions occurs via the ionization of H$_{2}$ by 
cosmic rays followed by electron excitation of H$_{2}$ leading to photon emission \citep{Gredel1989}.
We start from bare grain surfaces and use the same initial gas phase abundances as in~\citet{Chang2014}, which are based on the assumption of low-metal abundances.   
We run each model 25 times up to a time of $2 \times 10^{5}$ yr and take the average of these 25 simulation results so that
the minimum non-zero fractional abundance of each species in the model is $4\times 10^{-13}$.

\subsubsection{New Gas Phase Reactions}

The more important additional gas-phase chemical reactions are listed in Table~\ref{table1}; 
of these, a few  have been studied at 10 K, while some of the reaction rate coefficients have been extrapolated from higher 
 temperature values, and still others represent a higher degree of speculation. 
 The important formation of methoxy in the gas-phase  via the reaction between OH radical and methanol:
\begin{equation}
\label{OHCH3OH}
{\rm OH +  CH_{3}OH \longrightarrow CH_{3}O + H_{2}O}
\end{equation}
is well studied \citep{Shannon2013}. 
This reaction  proceeds via an intermediate metastable van der Vaals-type complex, CH$_3$OHOH$^*$, which can 
redissociate into reactants or proceed to form the final products via tunneling.
 
Contrary to our earlier work~\citep{Chang2014}, we  distinguish the methoxy and hydroxymethyl (CH$_2$OH) radicals in this study.  
We include the new gas-phase reactions related to methoxy, dimethyl ether and methyl formate  in~\citet{Vasyunin2013a} 
and \citet{Balucani2015} except chemical reactions involving atomic fluorine because F can be quickly locked in the form of HF, so that  Cl atoms play 
the major role in converting CH$_3$OCH$_3$ to CH$_3$OCH$_2$ in the gas phase~\citep{Balucani2015}.  
Moreover, the additional gas-phase loss pathways of CH$_3$O, H$_2$CO, CH$_3$OCH$_2$, CH$_3$CHO and CH$_2$OH
in ~\citet{Ruaud2015} are also included in our reaction network.
Because CH$_3$O and CH$_3$OCH$_2$ are crucial for the gas phase synthesis of COMs, we
included destruction reactions of these two species by ions to complete the gas phase
destruction pathways. The rate coefficients and products 
of ion - CH$_3$O reactions are set to be the same as those for ion -CH$_2$OH reactions, 
which already exist in our gas-grain network. Similarly, the rate coefficients of ion - CH$_3$OCH$_2$ reactions
are set to be the same as those of ion - CH$_3$OCH$_3$ reactions in our original gas grain network.
We also introduced the channel CH$_3$OCH$_3$$^{+}$  +  e$^{-}$ $\rightarrow$ CH$_3$OCH$_2$  +  H
to complete the gas-phase formation routes of CH$_3$OCH$_2$,
and we have included the gas-phase, mainly destruction,  reactions related to glycolaldehyde (CH$_2$OHCHO) \citep{Garrod2008,Belloche2014,Garrod2015}.

\citet{Shannon2014} have shown that atomic C can react quickly with methanol  at low temperatures and
this reaction may be the major loss process in the gas phase when atomic C is abundant.  We included
this reaction in our reaction network.  In addition, we included  
three reactions involving the radical CH and the species H$_{2}$O, CH$_{3}$OH, and H$_{2}$CO; these are analogs of surface reactions 
which are shown in Table~\ref{table20}.

\begin{table}
\caption{Some Additional Gas-Phase Reactions }
\label{table1}
\begin{tabular}{lllll}
  \hline
Additional Reaction &     &                                               & Rate Coefficient &    References \\
             &     &                                              & at 10 K(cm$^3$ s$^{-1}$)                  &               \\
             \hline
(1) OH + CH$_3$OH & $\rightarrow$ & CH$_3$O + H$_2$O               & 3.0$\times$10$^{-10}$     & \citet{Shannon2013} \\
(2) CH$_3$ + CH$_3$O & $\rightarrow$ & CH$_3$OCH$_3$ + photon     & 3.0$\times$10$^{-10}$     & \citet{Balucani2015}\\
(3) CH$_3$OCH$_2$ + O & $\rightarrow$ & HCOOCH$_3$ + H             & 2.0$\times$10$^{-10}$     & \citet{Hoyermann1996} \\
(4) CH$_3$OCH$_3$ + Cl & $\rightarrow$ & CH$_3$OCH$_2$ + HCl       & 2.0$\times$10$^{-10}$     & \citet{Wallington1988} \\
(5) CH$_3$O + O    & $\rightarrow$ &  CH$_{3}$ + O$_{2}$                  & 2.5$\times$10$^{-11}$     & \citet{Zellner1987}      \\
(6) CH$_3$O + H    & $\rightarrow$ & H$_2$CO + H$_2$               & 3.0$\times$10$^{-11}$     & \citet{Baulch1992}    \\
(7) CH$_3$O + C    & $\rightarrow$ & CH$_3$ + CO                   & 3.0$\times$10$^{-10}$     & \citet{Ruaud2015}\\
(8) CH$_3$O + N   & $\rightarrow$ & H$_2$CO + NH                  & 5.6$\times$10$^{-12}$     & \citet{Ruaud2015}\\
                  &  $\rightarrow$ & CH$_3$ + NO                   & 1.7$\times$10$^{-12}$     &                   \\
(9) CH$_3$OCH$_2$ + H & $\rightarrow$ & CH$_3$O + CH$_3$           & 3.0$\times$10$^{-11}$    & \citet{Ruaud2015}\\
(10) CH$_3$OCH$_2$ + N & $\rightarrow$ & CH$_3$O + H$_2$CN         & 3.0$\times$10$^{-11}$    & \citet{Ruaud2015}\\
(11) CH$_3$OCH$_2$ + C & $\rightarrow$ & CH$_3$O + C$_2$H$_2$      & 3.0$\times$10$^{-10}$    & \citet{Ruaud2015}\\      
(12) CH$_3$CHO  + C    & $\rightarrow$ & C$_2$H$_4$ + CO           & 3.0$\times$10$^{-10}$     & \citet{Ruaud2015}\\ 
(13) CH$_2$OH + H & $\rightarrow$ & H$_2$CO + H$_2$                & 1.0$\times$10$^{-11}$     & \citet{Ruaud2015}\\
                  & $\rightarrow$ & CH$_3$ + OH                    & 1.6$\times$10$^{-10}$     & \citet{Ruaud2015}\\
(14) CH$_2$OH + O & $\rightarrow$ & H$_2$CO + H$_2$                & 1.0$\times$10$^{-10}$     & \citet{Ruaud2015}\\
(15) CH$_2$OH + C & $\rightarrow$ & CH$_3$ + CO                    & 3.0$\times$10$^{-10}$     & \citet{Ruaud2015}\\
(16) CH$_2$OH + N & $\rightarrow$ & H$_2$CO + NH                   & 2.2$\times$10$^{-11}$     & \citet{Ruaud2015}\\
                  & $\rightarrow$ & HCN + H$_2$O                   & 3.4$\times$10$^{-11}$     & \citet{Ruaud2015}\\
(17) H$_2$CO + C & $\rightarrow$ & CH$_2$ + CO                     & 3.0$\times$10$^{-10}$    & \citet{Ruaud2015}\\
(18)  H$_2$CO + CH & $\rightarrow$ & CH$_3$ + CO                   & 4.0$\times$10$^{-10}$    & \citet{Ruaud2015}\\
(19) H$_2$CO + CN & $\rightarrow$ & HCN + HCO                      & 3.9$\times$10$^{-11}$     & \citet{Ruaud2015}\\   
(20) CH$_3$OCH$_3^{+}$ + e$^{-}$ & $\rightarrow$ & CH$_3$OCH$_2$ + H  & 2.7$\times$10$^{-8}$   & this work \\
(21) CH  +   H$_2$O  & $\rightarrow$ & H$_2$CO  +  H & 5.6$\times$10$^{-10}$  &  \citet{Bergeat2009}  \\
(22) CH  +   CH$_3$OH & $\rightarrow$ & H$_2$CO  +  CH$_3$  &  2.5$\times$10$^{-10}$  &  \citet{Ruaud2015} \\ 
(23) CH  +   H$_2$CO  &  $\rightarrow$ & CO  +  CH$_3$ &  4.0$\times$10$^{-10}$    &  \citet{Ruaud2015} \\
(24) C  +   CH$_3$OH  &  $\rightarrow$ & HCO  +  CH$_3$  &  2.0$\times$10$^{-10}$     &  \citet{Shannon2014}  \\
(25) CH$_3$O + ions &               &                               &                           & this work \\
      e.g.          &               &                               &                           &          \\
      H$_3^{+}$ + CH$_3$O  &$\rightarrow$ & CH$_3$OH + H$_2$    &   2.2$\times$10$^{-8}$       &           \\
      He$^{+}$  +   CH$_3$O &$\rightarrow$ & CH$_2^{+}$ + OH + He & 9.3$\times$10$^{-9}$       &           \\
(26) CH$_3$OCH$_2$ + ions &         &                               &                           & this work \\
      e.g.                &          &                              &                           &          \\
     H$_3^{+}$ + CH$_3$OCH$_2$ &$\rightarrow$ & CH$_3$OCH$_3^{+}$ + H$_2$ & 1.6$\times$10$^{-8}$ &           \\
     He$^{+}$ + CH$_3$OCH$_2$ &$\rightarrow$ & CH$_3^{+}$ + H$_2$CO + He &7.2$\times$10$^{-9}$ &           \\
(27) glycolaldehyde   &     &                            &                           & \citet{Garrod2015} \\
     reactions, e.g.                &          &                              &                           &          \\
     H$_3^{+}$ + CH$_2$OHCHO &$\rightarrow$ & CH$_2$OHCH$_2$O$^{+}$ + H$_2$ & 3.4$\times$10$^{-8}$ &      \\
     He$^{+}$ + CH$_2$OHCHO &$\rightarrow$ & CH$_2$OH +  HCO$^{+}$ +  He    & 1.5$\times$10$^{-8}$ &      \\\hline               
\end{tabular}
\end{table}

\begin{table}
\caption{ Some Additional Surface Reactions }
\label{table20}
\begin{tabular}{llll}
  \hline
  (1) CH  +   H$_2$O  &  $\rightarrow$ &  H$_2$CO  +  H & \citet{Bergeat2009}  \\
 (2) CH  +   CH$_3$OH  &  $\rightarrow$ & H$_2$CO  +  CH$_3$ & \citet{Ruaud2015} \\ 
 (3) CH  +   H$_2$CO  &  $\rightarrow$ &  CO  +  CH$_3$ &  \citet{Ruaud2015} \\
 (4) OH  +   CH$_3$OH  & $\rightarrow$ &  H$_2$O  +  CH$_3$O  &  \citet{Shannon2013} \\ 
                             & $\rightarrow$ &  H$_2$O  +  CH$_2$OH  &  \citet{Mccaulley1989} \\
\end{tabular}
\end{table}

\subsubsection{Aspects of COM Grain Chemistry}

 The methoxy and hydroxymethyl radicals lead to COMs on grain surfaces.  Both
 radicals can be hydrogenated to produce methanol while only methoxy can react with HCO and CH$_3$ 
to form HCOOCH$_3$ and CH$_3$OCH$_3$ respectively, while hydroxymethyl can react with HCO and CH$_3$ to form
glycolaldehyde (CH$_{2}$OHCHO) and ethanol (C$_2$H$_5$OH) instead.  As for their formation, 
accretion of methoxy onto the grains is not likely to be important because the fractional abundances of
gas-phase methoxy are found to quite low in the cold cores B1-b and L1544~\citep{Bacmann2012,Cernicharo2012,Vastel2014}.
On the grain surface, both methoxy and hydroxymethyl can be produced via the photodissociation of methanol \citep{Garrod2008}.
The hydrogenation of CO can lead to hydroxymethyl and methoxy.   
In most models we assume that Y, the methoxy yield by gradual hydrogenation of CO, is zero because 
it has been argued that the abundance of methoxy on grain surface is small and methoxy can quickly isomerize to
hydroxymethyl~\citep{Cernicharo2012}.  We do, however,  consider the possibility that Y=0.5 in one model, 
considering that the H + H$_{2}$CO reaction in the gas phase possesses a lower barrier 
to the formation of CH$_{3}$O than to CH$_2$OH, coupled with the high barrier to isomerization in the gas.
We also include the analog of reaction~(\ref{OHCH3OH})  
on the surfaces of grains among the new surface reactions in our model, shown in Table~\ref{table20}. 
If efficient, this reaction can be the dominant production mechanism for surface methoxy at certain times. It likely competes with stabilization to form the van der Waals complex CH$_{3}$OHOH \citep{Ruaud2015}.
 Moreover, we also consider the case that the product of reaction~(\ref{OHCH3OH}) is CH$_2$OH instead of CH$_3$O due to the high exothermicity 
of the chain reaction mechanism.  
The radical CH is important for the formation of 
COMs on grain surface, as explained in the next section, so we included three more surface reactions 
that destroy CH which are shown in Table~\ref{table20}.

\begin{table}
\caption{Model Parameters }
\label{table2}
\begin{tabular}{lllllllllll}
  \hline
Model & M40 & M41 & M42 & M43 & M02 & M12 & M22 & M32 & M431 & M432\\
\hline
$P$   & 0.1 & 0.1 & 0.1 & 0.1 & 0  & 0.01  & 0.03 & 0.05 &  0.1 &  0.1 \\ \hline
$RP$  & 0.0001 & 0.1 &0.5 &1.0 &0.5 & 0.5 & 0.5  & 0.5   &  1.0 &  1.0  \\ \hline
$Y$   &  0 &   0  &  0  &  0  &  0 &  0   &   0   &   0 &   0  &   0.5 \\ \hline
$RP_2$ &  0 &  0  &  0  &  0  &  0 &  0   &   0   &   0 &  1.0 &   1.0  \\ 
\hline
\end{tabular}
\tablecomments{
 The letter $P$ refers to reactive desorption efficiency, 
while $RP$ refers to the probability  that the surface reaction between 
OH and CH$_3$OH produces the methoxy or  hydroxymethyl radical as a product.
 The probability  that the surface reaction between
OH and CH$_3$OH produces the hydroxymethyl radical is represented as $RP_2$.
The letter $Y$ refers to the CH$_3$O yield by the hydrogenation of H$_2$CO.
See the text for details.}

\end{table}

We define a parameter, $RP$, for the probability that OH and CH$_3$OH,  with the radical lying atop methanol, complete a chain reaction to 
to form CH$_3$O and H$_2$O  or CH$_2$OH and H$_2$O. The larger the value of $RP$, the more likely OH can react with CH$_3$OH in the same site. 
In our simulations, we utilize values for $RP$ ranging from 10$^{-4}$ to unity. The parameter $RP_2$ is used to refer to
the probability that OH and CH$_3$OH form CH$_2$OH and H$_2$O. 
Table~\ref{table2} shows the values of the parameters  $P$, $RP$, $Y$ and $RP_2$ used for different models simulated in this work.  
The models are characterized by two  or three numbers following the letter M. For models with two numbers,    
the first number refers to the value of the efficiency $P$ of reactive desorption; 
the digits 0,1,2,3,4 refer respectively to values for $P$ of 0, 0.01, 0.03, 0.05 and 0.1.  
The second number refers to the probability $RP$.  Here the digits 0,1,2,3 refer respectively to 
values of $RP$ equal to 10$^{-4}$, 0.1, 0.5, and 1.0.  
For example, the model M40 refers to a simulation in which the probability of reactive desorption of products 
is 0.1 while the probability that methoxy and water are formed in the reaction between OH and CH$_{3}$OH is a small 10$^{-4}$.
If a model has three numbers,  the first two numbers refer to the same parameters as those in a model with two numbers,
while the third number is an index that specifies different $Y$ and $RP_2$ values. For instance, in model M431, $P$ = 0.1, $RP$ = 1.0,
$Y$=0 and $RP_2$=1.0. 

\section{Results: The Formation of COMs}
In this section, we discuss how COMs are formed both in the gas and on grains followed by desorption into the gas, in an attempt to determine which mechanism, if either,  is dominant for which molecule.  We start with a discussion of the uncertain abundance of the important precursor methoxy.

\subsection{ The Role and Abundance of Granular Methoxy}

\begin{figure}
\centering
\resizebox{15cm}{15cm}{\includegraphics{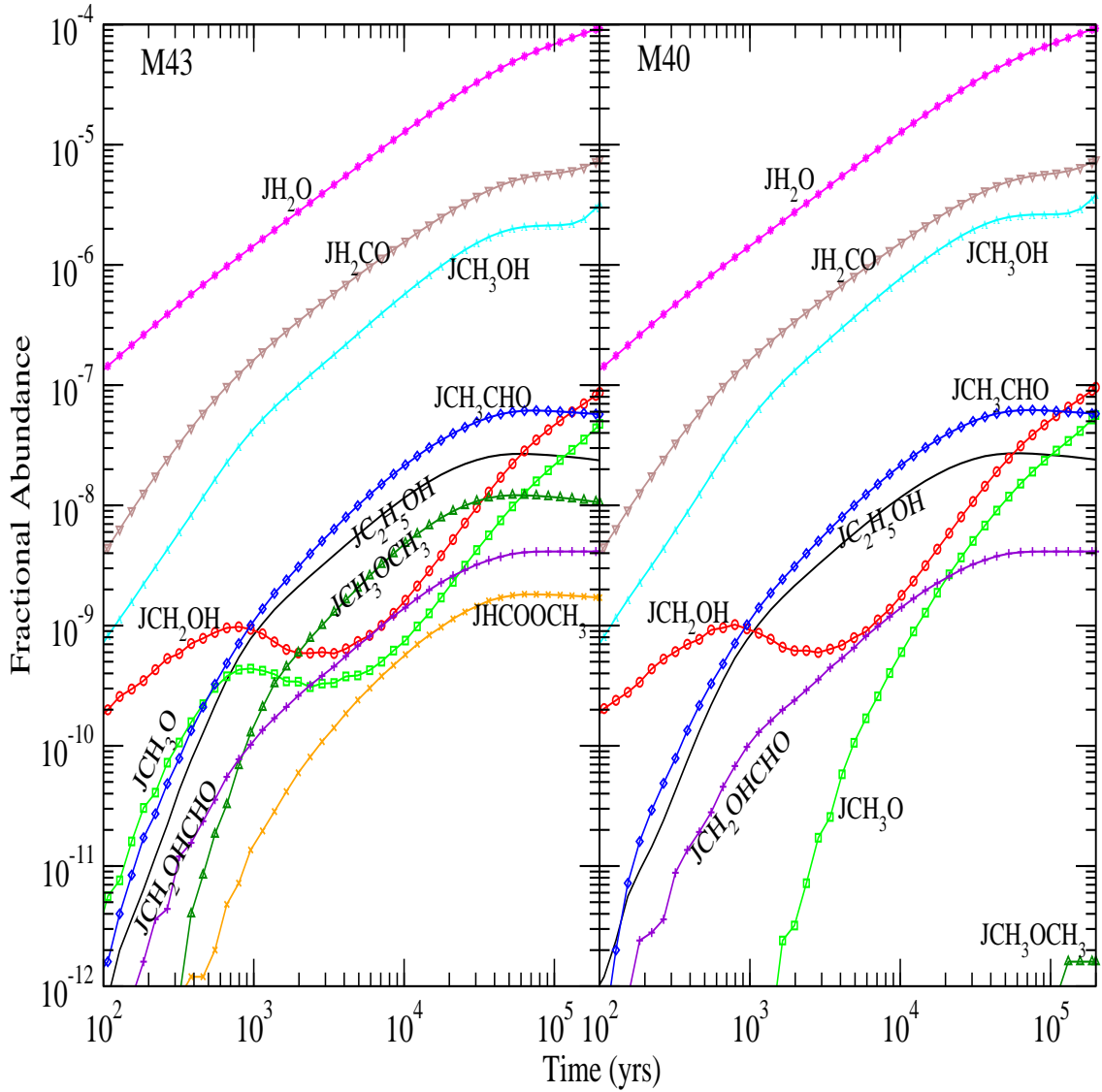}}
\caption{ The fractional abundances of COMs, methoxy, hydroxymethyl and water on and in grain mantles 
as a function of time in models M43 ($P = 0.1$, $RP = 1.0$, $Y=0$, $RP_2=0$) and M40 ($P = 0.1$, $RP = 10^{-4}$,  $Y=0$, $RP_2=0$). 
}
\label{fig1}
\end{figure}

 Fig. \ref{fig1} shows the fractional abundances with respect to  $n_{\rm H}$ for
HCOOCH$_3$, CH$_2$OHCHO, CH$_3$OCH$_3$,  C$_2$H$_5$OH, CH$_3$CHO, CH$_3$O, CH$_2$OH, H$_2$CO and CH$_3$OH as  functions of time
on dust grain ices in models M43 ($P = 0.1$, $RP = 1.0$, $Y=0$, $RP_2=0$) and M40 ($P = 0.1$, $RP = 10^{-4}$, $Y=0$, $RP_2=0$).  
In the figure, the letter J designates granular species.
These are extreme cases; the models in which the parameter $RP$ assumes intermediate values have results in between M43 and M40.
We first discuss the  abundance of methoxy  because its existence on ice mantles to any appreciable degree is now 
controversial~\citep{Cernicharo2012}.
In model M43, methoxy can be efficiently formed on grain surfaces by the surface analog of reaction~(\ref{OHCH3OH})
while in model M40, this reaction contributes little to the formation of methoxy on grain surfaces.
Reaction~(\ref{OHCH3OH}) cannot happen on grain surfaces 
by a diffusive process at 10 K because neither OH nor
CH$_3$OH can diffuse appreciably on grain surfaces this cold. Instead, we invoke a chain reaction mechanism, in which the reaction happens
when O combines with atomic H and there is an CH$_3$OH molecule lying below the newly formed OH so that OH can immediately
react with CH$_3$OH to form CH$_3$O and water.

In model M43, the methoxy abundance first increases because of the increase in the granular methanol abundance via hydrogenation of CO followed by surface reaction~(\ref{OHCH3OH}), then
fluctuates before $10^4$ yr, and then increases again because photodissociation of methanol can produce methoxy, which is buried within the  ice  mantle. 
In  model M40, on the other hand,  there is hardly any methoxy on the grain surface before $2\times10^3$ yr
because methoxy can only be formed weakly by surface reaction~(\ref{OHCH3OH}). Photodissociation of methanol is the major
reaction to form methoxy in model M40, and at later times this mechanism produces as much methoxy as in Model M43. 
Indeed, after $2\times10^3$ yr
the abundance of surface methoxy in model M40 increases monotonically. 
It is interesting to see that at about $10^4$ yr, the total (surface +bulk) abundance of methoxy in both models M40 and M43 is roughly the same because methoxy formed by photodissociation of methanol that is deeply buried within ice accumulates and is much more abundant than
methoxy on the topmost ice layer.  Nevertheless, the abundance of methoxy on the topmost layer of ice mantle in M43 is 
always much larger than that in M40 because the surface reaction~(\ref{OHCH3OH}) is efficient and is the major pathway for the formation of CH$_3$O on the topmost layer in M43.  Moreover,  it is the methoxy on the topmost layer
that can react with radicals to form methyl formate and dimethyl ether via the chain reaction mechanism on grain surface.
Methoxy buried in the ice mantle can only participate in the interstitial chemistry or photodissociation.  
  
The abundance of surface methoxy can be much lower than calculated here if, as discussed by  \citet{Cernicharo2012}, laboratory evidence suggests that surface methoxy can isomerize to CH$_{2}$OH 
if it does not react with H first to form methanol.  The main piece of evidence for this isomerization is the inability to 
detect the infrared bands of methoxy.  Although the authors suggest that the isomerization is promoted by proton exchange, 
it is not clear to us how this process can occur quickly on an interstellar ice surface, even if it does in the laboratory. 
We have decided not to include this isomerization  until it is investigated more closely. 
Calculations of other ion-neutral processes on ice mantles have been undertaken by \citet{Woon2011}.
Perhaps some information on the isomerization could also be obtained from measured abundances of CH$_{3}$OD and CH$_{2}$DOH 
on interstellar grains.  On the other hand, our assumption that methoxy cannot be formed by gradual hydrogenation of CO 
can partially offset the absence of isomerization of methoxy in our surface reaction network. 

\begin{figure}
\centering
\resizebox{15cm}{15cm}{\includegraphics{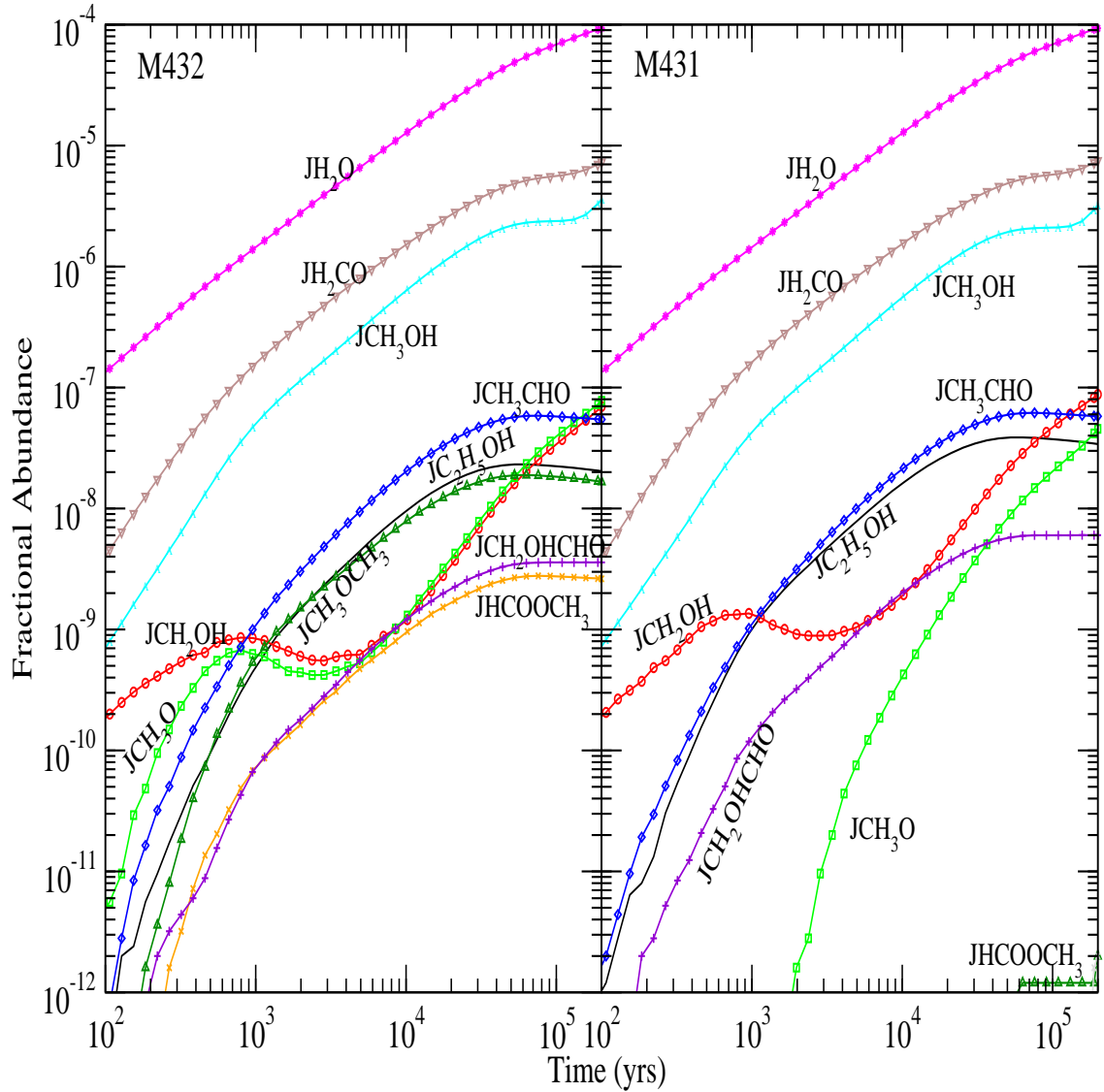}}
\caption{ The fractional abundances of COMs, methoxy, hydroxymethyl and water on and in grain mantles 
as a function of time in models M431 ($P = 0.1$, $RP = 1.0$, $Y=0$, $RP_2=1.0$) and M432 ($P = 0.1$, $RP = 1.0$, $Y=0.5$, $RP_2=1.0$). 
}
\label{fig2}
\end{figure}

Fig. \ref{fig2} shows the fractional abundances with respect to  $n_{\rm H}$ for surface
HCOOCH$_3$, CH$_2$OHCHO, CH$_3$OCH$_3$,  C$_2$H$_5$OH, CH$_3$CHO, CH$_3$O, CH$_2$OH, H$_2$CO and CH$_3$OH as functions of time
on dust grain ices in model M431 where the product of OH and CH$_3$OH is CH$_2$OH instead of CH$_3$O and model M432 where
the CH$_3$O yield of the H$_2$CO hydrogenation reaction is 0.5. 
The reactive desorption efficiency is also fixed to be 0.1 in models M431 and M432.
We can see that, similar to model M40, 
there is hardly any surface methoxy formed in model M431 before $2\times 10^3$ yrs because there is 
no efficient methoxy formation process on the topmost layer in model M431. However, if the CH$_3$O yield of the H$_2$CO hydrogenation 
reaction is non-negligible, specifically $Y= 0.5$ as in model M432, the surface methoxy abundance in M432
is even slightly higher than that in model M43 before $2\times 10^3$ yrs because H$_2$CO 
hydrogenation is an efficient surface CH$_3$O formation process on the topmost layer. 
After $10^4$ yrs, however, as the accumulation of CH$_3$O formed by photodissociation reactions becomes
the major source of CH$_3$O on grain surfaces, methoxy abundances in 
models M432 and M431 are almost the same.

\subsection{The Production of COMs on Ice}
While the larger COMs cannot be detected on granular ices, it is important to consider their abundance prior to desorption into the gas, if it can occur. We can see that the abundances of 
water and methanol, as shown in Fig.~\ref{fig1},  always increase with time in both models M43 and 
M40 because they are continuously formed on dust grains. Since the  surface reaction~(\ref{OHCH3OH})
is not the major reaction to form or destroy water or methanol, the abundances of water and methanol do not change 
with the change in efficiency of this reaction.

\begin{figure}
\centering
\resizebox{15cm}{15cm}{\includegraphics{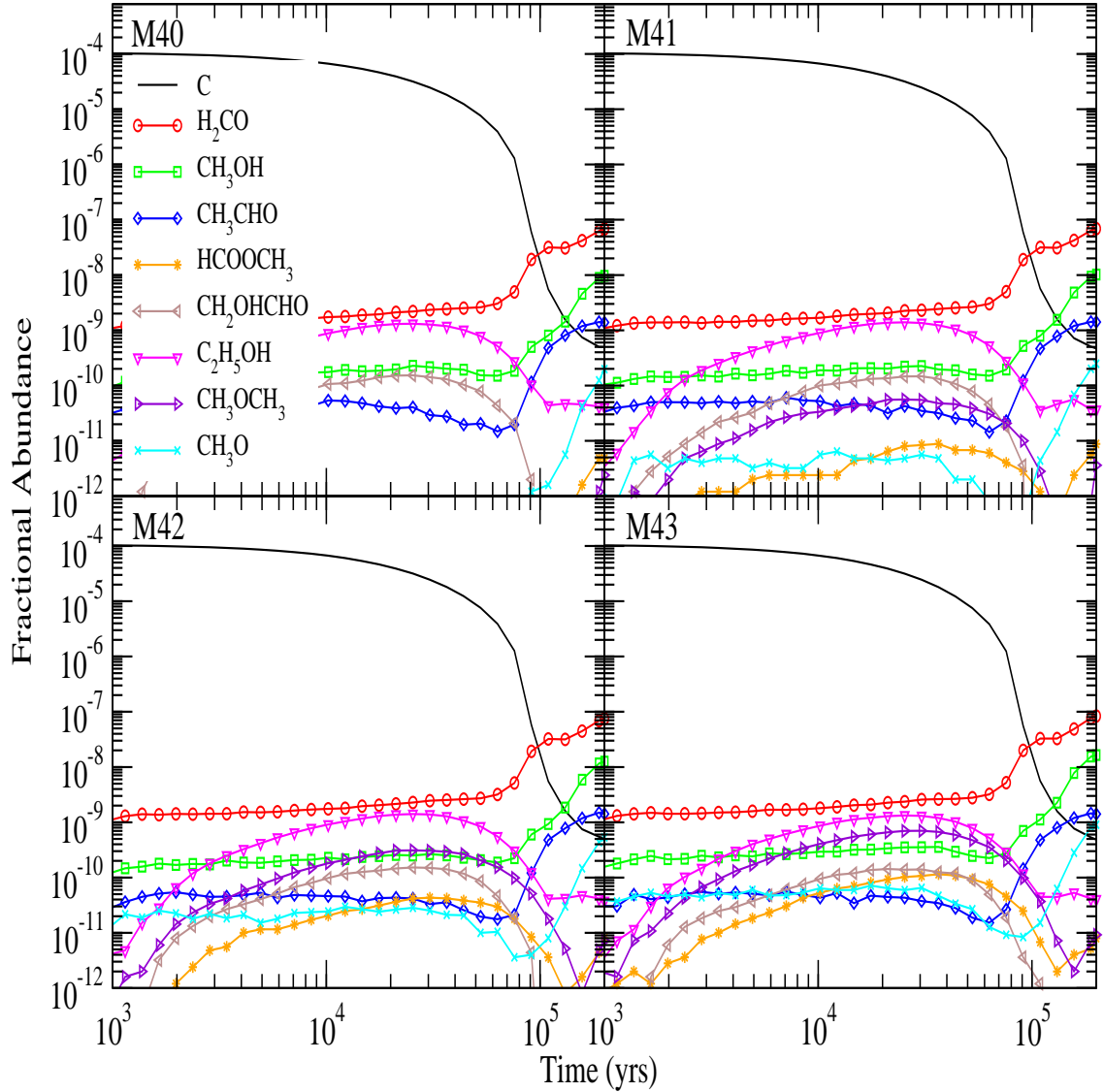}}
\caption{ The fractional abundances of gas phase COMs, methoxy, and atomic carbon as a function of time for models M40, M41, M42 and M43.  See Table 2 for a definition of the models.
}
\label{fig3}
\end{figure}

The abundances of methyl formate, dimethyl ether, glycolaldehyde, ethanol, and acetaldehyde first increase with time and then 
reach a plateau at $10^5$ yr in both models M43 and M40. Similarly, although these five COMs can be formed by the 
recombination of radicals in our surface reaction network, 
they cannot be formed on grain surfaces by a diffusive process at 10 K because relevant radicals
cannot diffuse appreciably on grain surfaces this cold.  
Instead, they can be formed by similar chain reaction mechanisms.
Acetaldehyde is produced  when CH$_2$ combines with atomic H and there is an HCO radical lying below the newly formed CH$_3$. 
Methyl formate can be produced by a newly formed HCO, which is formed by CH and O, and a CH$_3$O molecule
 lying below the  HCO,  
while  dimethyl ether  can  be produced by a newly formed CH$_3$ radical and a CH$_3$O
radical  lying below it. The isomer CH$_2$OH can combine with HCO  or CH$_3$ to form glycolaldehyde or ethanol
in a similar way to the formation of methyl formate or dimethyl ether.  
The radical CH$_2$OH itself can be formed by hydrogenation of H$_2$CO 
and photodissociation of methanol, so the abundance of CH$_2$OH as a function of time in both models M43 and M40 is similar to
that of CH$_3$O in model M43, where the surface reaction~(\ref{OHCH3OH}) is  efficient.   
We can also see that because the formation pathways of glycolaldehyde and acetaldehyde are independent of
the radical CH$_3$O on the topmost layer, the abundances of glycolaldehyde and acetaldehyde are almost independent of the value of $RP.$ 
On the other hand, because the formation mechanisms of methyl formate and dimethyl ether are dependent on 
the radical CH$_3$O on the topmost layer, the abundances of dimethyl ether and methyl formate in model  M40 are more than two 
orders of magnitude smaller than in model M43.
The plateaus of methyl formate, 
dimethyl ether, glycolaldehyde, ethanol, and acetaldehyde abundances ,  reached after $10^5$ yr, occur because gas phase atomic carbon is depleted
 after $10^5$ years as shown in Fig.~\ref{fig3} while surface CH and CH$_2$ 
are mainly formed by hydrogenation of accreted C from the gas phase.

\begin{figure}
\centering
\resizebox{15cm}{15cm}{\includegraphics{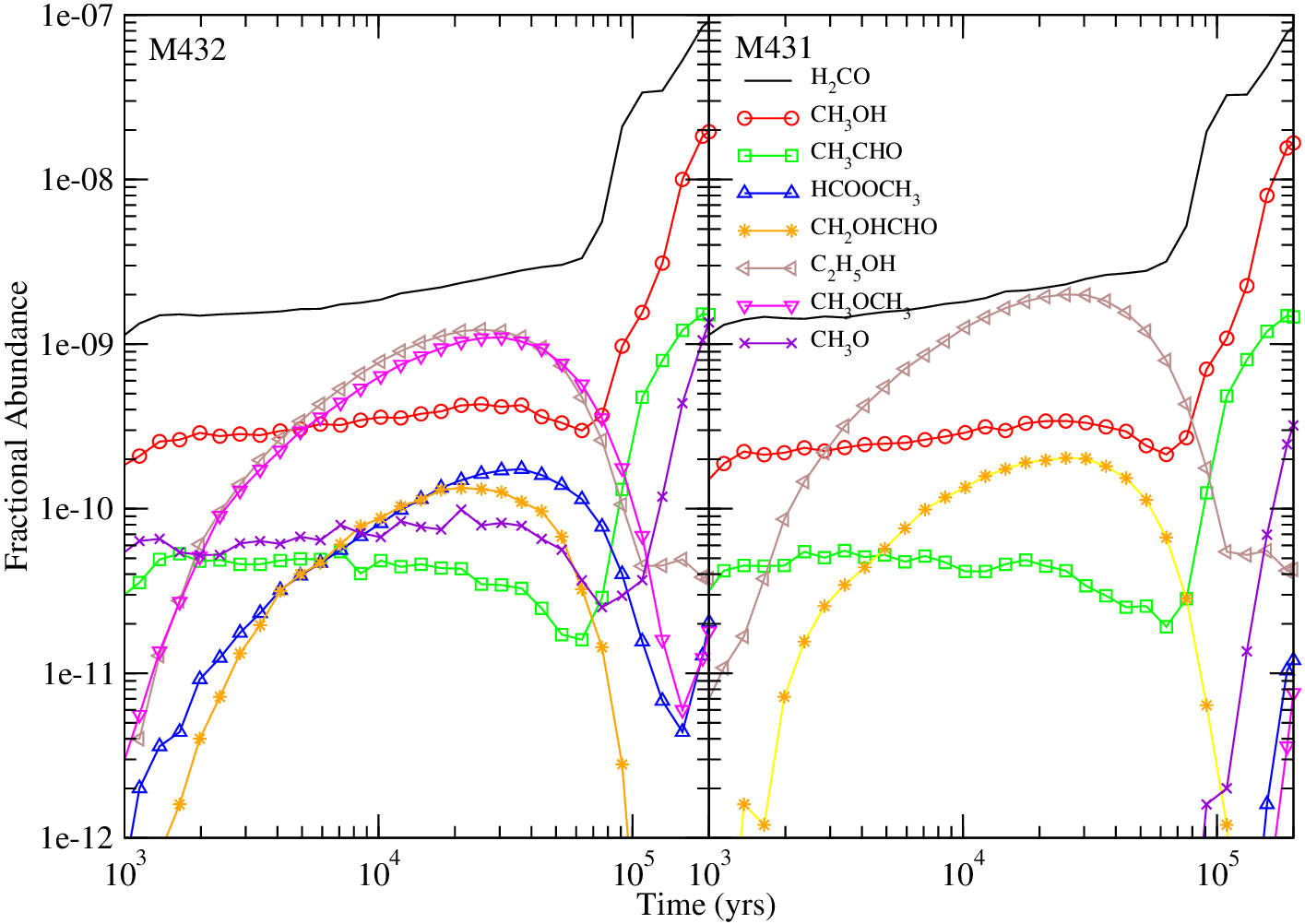}}
\caption{ The fractional abundances of gas phase COMs, methoxy, and atomic carbon as a function of time for models M431 and M432.  See Table 2 for a definition of the models.
}
\label{fig4}
\end{figure}

 Fig.~\ref{fig2} shows that if the product of surface reaction~(\ref{OHCH3OH}) is CH$_2$OH instead of CH$_3$O, methyl formate  or dimethyl ether 
can hardly be formed on grain surfaces in model M431 because only a few methoxy molecules can be formed on the topmost layer of ice mantle. Moreover,
if the CH$_3$O yield of H$_2$CO hydrogenation takes the non-negligible value of 0.5, a moderate amount of methyl formate and dimethyl ether
can be formed on grain surfaces in model M432 even if the surface reaction~(\ref{OHCH3OH}) produces CH$_2$OH instead of CH$_3$O.
On the other hand, the abundances of surface species other than methyl formate, dimethyl ether and methoxy in 
Fig.~\ref{fig2} are not much affected
by the change of surface reaction~(\ref{OHCH3OH}) products and the CH$_3$O yield of H$_2$CO hydrogenation because the formation of these species
is not strongly dependent on the methoxy abundance on the topmost layer. The abundance of surface CH$_2$OH before $2\times 10^5$ yrs in model M431
is less than 2 times more than  that in model M432 because the CH$_2$OH yield of H$_2$CO hydrogenation in model M431 is 2 times as much as
that in model M432. Consequently, the abundances of surface species such as C$_2$H$_5$OH, which require CH$_2$OH as a precursor in model M431,  
is about a factor of 2 larger than that in model M432.

\subsection{The Production of Gaseous COMs}

Gas-phase COMs can be produced by gas-phase processes, granular processes followed by non-thermal desorption, or both of the above either in parallel or in series.  
The fractional abundances of gas-phase COMs, methoxy, and atomic carbon as a function of time for models M40, M41, M42 and M43 
are presented in Fig. \ref{fig3}. In these models, the efficiency of reactive desorption remains at 0.1, 
while the efficiency of methoxy production in surface reaction~(\ref{OHCH3OH}) increases from near zero to unity.  
Similar to the abundances of surface species, the abundances of gas phase carbon atoms, formaldehyde,
methanol, acetaldehyde,  ethanol and glycolaldehyde are independent of the parameter $RP$ while the abundances of methoxy, methyl formate 
and dimethyl ether typically increase  as $RP$ becomes larger. Methanol can be be continuously produced on grain surfaces via the  
 hydrogenation of CO, and then be desorbed efficiently into the gas phase; however, because gaseous methanol can be destroyed 
by gas phase C atoms, its abundance maintains only a moderate value and then quickly increases as gaseous C is depleted in all models.
Glycolaldehyde is another species that is mainly formed on the grain surface. However, the formation of glycolaldehyde is strongly 
dependent on the abundance of the surface radical CH, which is mainly produced by hydrogenation of atomic C accreted
on grain surfaces.  So after $10^5$ yr, when the abundance of gas phase C quickly drops, the formation rate of glycolaldehyde 
on the grain surface also quickly drops; thus, the abundance of gas phase glycolaldehyde also drops quickly.  
As in the formation of glycolaldehyde, the formation of
methyl formate and dimethyl ether on grain surfaces also depend on accreted gas phase C; however, because 
these species can also be synthesized in the gas phase,  their abundances
first drop as gas phase C is depleted and then increase as the abundance of gas phase CH$_3$O, CH$_3$OH and H$_2$CO increase. 
Similarly,  the gas phase abundance of gaseous ethanol also drops at about $10^5$ yrs as gas phase C is depleted.
In addition to the grain surface formation route, ethanol can  be formed by dissociative recombination of
C$_2$H$_5$OH$_2^{+}$, which is produced by  two gas phase radiative association reactions in our gas-grain network:
\begin{equation}
{\rm H_3O^{+} + C_2H_4 \longrightarrow C_2H_5OH_2^{+} },
\end{equation}
\begin{equation}
{\rm H_2O + C_2H_5^{+} \longrightarrow C_2H_5OH_2^{+} }.
\end{equation}
So, after gaseous C is depleted, unlike glycolaldehyde, the abundance of gaseous ethanol still maintain a small value 
(a few $\times$ $10^{-11}$) because of its gas phase synthetic routes.

From Figure~\ref{fig3}, we can also see that gaseous
methyl formate  and dimethyl ether have very low abundances in model M40. 
As $RP$ increases, desorbed dimethyl ether and methyl formate molecules 
become the major source of gas-phase dimethyl ether and methyl formate. 
In addition, the gas phase synthetic routes CH$_3$ + CH$_3$O $\rightarrow$ CH$_3$OCH$_3$ and O + CH$_3$OCH$_2$ $\rightarrow$ H + HCOOCH$_3$
also become more significant as $RP$ increases because the abundances of gas phase CH$_3$O and CH$_3$OCH$_3$ increase. 

In models M41, M42 and M43
methoxy can be produced in both the gas phase and on the grain surface while in M40, it is mainly formed in the gas phase. The increase
of gas phase methoxy after $10^5$ yr occurs because gas phase species that can react with CH$_3$O such as O, N and C are depleted
and because its precursor, methanol increases in abundance. 
Acetaldehyde is a COM that can also be synthesized both on dust grains 
and in the gas phase. The abundance of gas phase C also must be high enough in order to produce CH$_3$CHO on dust grains 
as explained above; however, since CH$_3$CHO can also be produced by the reaction 
\begin{equation}
{\rm O + C_{2}H_{5} \longrightarrow   CH_{3}CHO + H ,}
\end{equation}
in the gas phase where C$_2$H$_5$ is desorbed
into the gas phase by reactive desorption, the abundance of gas phase CH$_3$CHO is significant at all times in models M40, M41, M42 and M43.
Table~\ref{table3} lists the dominant synthetic routes for gaseous COMs - gas-phase or grain-surface -  for Models M40 M41, M42, and M43  
at a time of $5 \times 10^4$ yr. The results show that with an efficient chemical desorption mechanism ($P = 0.1$), 
synthesis occurs mainly on grains with the exception of acetaldehyde.

 Similar to model M40, as shown in Fig.~\ref{fig4}, gaseous methyl formate and dimethyl ether in model M431 have very low abundances
because the abundance of methoxy on the topmost layer on ice mantle is very low.  Thus,  it is difficult to synthesize methyl formate and dimethyl ether
on grain surfacea. After we introduce an efficient surface CH$_3$O formation process, i.e. hydrogenation of H$_2$CO, 
in model M432, the abundances of gaseous methyl formate and dimethyl ether are similar to that in model M43. Generally speaking,
the abundances of gas phase species other than methoxy, methyl formate, and dimethyl ether do not vary much in models
M40, M41, M42, M43, M431 and M432. We can also see that the time dependence of gas phase species in Models M432 amd M431 is similar 
to that in models M43 and M40 respectively.

\begin{table}
\caption{Dominant Synthetic Mechanism for Gaseous COMs. I.}
\label{table3}
\begin{tabular}{lllllll}
  \hline
Model         &  M40    & M41   & M42   & M43  & M431 & M432\\
\hline
H$_2$CO       & Both    & Both  & Both  & Both  &  Both &  Both\\
CH$_3$OH      & Grain   & Grain & Grain & Grain  &  Grain &  Grain \\
HCOOCH$_3$    & --     & Grain & Grain & Grain  &  -- &  Grain\\
CH$_2$OHCHO   & Grain   & Grain & Grain & Grain  &  Grain &  Grain\\
CH$_3$OCH$_3$ &  -- &  Grain  & Grain & Grain &  -- &  Grain \\  
CH$_3$CHO     & Gas    & Gas    & Gas  & Gas   &  Gas &  Gas \\
C$_2$H$_5$OH  &  Grain  &  Grain  &  Grain &  Grain &  Grain &  Grain  \\
\end{tabular}
\tablecomments{
Calculations refer to a time of $5 \times 10^4$ yr, close to best fits in L1689b and B1-b.
Grain: gas phase COMs are formed mainly by surface reactions followed by reactive desorption.
Gas: gas phase COMs are formed mainly by gas phase reactions.
``--'': vanishingly small production.}
\end{table}

\begin{table}
\caption{Dominant Synthetic Mechanism for Gaseous COMs. II.}
\label{table4}
\begin{tabular}{lllll}
  \hline
Model         &  M02    & M12   & M22   & M32 \\
\hline
H$_2$CO       & Gas     & Both  & Both  & Both \\
CH$_3$OH      & Grain   & Grain & Grain & Grain \\
HCOOCH$_3$    & -- & Grain & Grain & Grain \\
CH$_2$OHCHO   & -- & Grain & Grain & Grain \\
CH$_3$OCH$_3$ & -- & Grain & Grain & Grain  \\  
CH$_3$CHO     & -- & Gas   & Gas   & Gas      \\
 C$_2$H$_5$OH  &  --  &  Grain  &  Grain &  Grain \\
\end{tabular}
\tablecomments{
Calculations refer to a time of $5 \times 10^4$ yr, close to best fits in L1689b and B1-b.
Grain: gas phase COMs are formed mainly by surface reactions followed by reactive desorption.
Gas: gas phase COMs are formed mainly by gas phase reactions.
``--'': vanishingly small production.}
\end{table}

\subsection{The Influence of Reactive Desorption}

\begin{figure}
\centering
\resizebox{15cm}{15cm}{\includegraphics{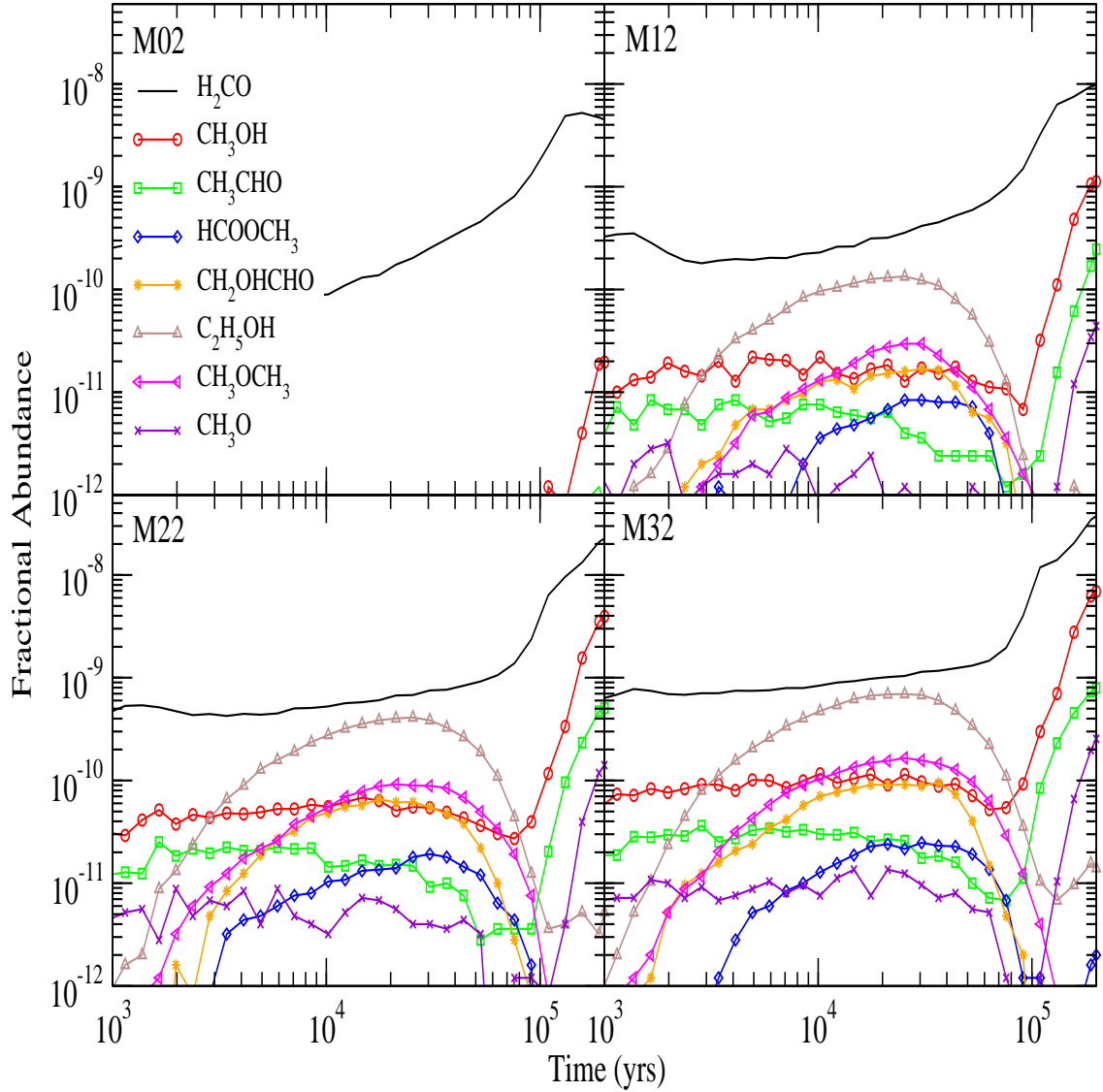}}
\caption{ The fractional abundances of gas phase COMs and methoxy as a function of time for models M02, M12, M22 and M32. See Table 2 for definitions of models.
}
\label{fig5}
\end{figure}

The fractional abundances of gas phase COMs and methoxy as a function of time for models M02, M12, M22 and M32
are presented in Fig.~\ref{fig5}.  In these 4 models, we fix $RP$ to be 0.5 
while $P$, the efficiency of reactive desorption,  takes the values 0, 0.01, 0.03 and 0.05  in models M02, M12, M22 and M32, respectively.
The products of surface reaction~(\ref{OHCH3OH}) are fixed to be CH$_3$O and H$_2$O, while hydrogenation of H$_2$CO can only lead to CH$_2$OH.
A  general feature that is apparent is that as the efficiency of reactive desorption goes from 0 to non-zero, 
the abundances of COMs become large enough to be displayed in the figure.  
With no reactive desorption,   although there is still an efficient methoxy formation pathway on the topmost layer, 
the abundance of gas phase COMs are very low because they or their precursors  cannot be synthesized 
efficiently in the gas phase with our reaction network. 
However, despite the fact that there is no efficient mechanism to desorb COMs from grain surfaces,
small amount of COMs, especially methanol and acetaldehyde, can be desorbed into the gas phase by photodesorption. In model
M12, where the reactive desorption efficiency is 0.01, the fractional abundance of ethanol gains the most, rising from a value 
below $10^{-12}$ to around $10^{-10}$, while the abundance 
of methanol increases by more than one order of magnitude.
With the same increase in reactive desorption efficiency, the fractional abundances of acetaldehyde and glycolaldehyde increase 
from a value below or slightly above $10^{-12}$ to a value that is typically around $10^{-11}$, 
while the abundances of methyl formate and dimethyl ether increase from a value
below $10^{-12}$ to a value that is around $10^{-11}$. 
The fractional abundance of methoxy increases from a value below 10$^{-12}$ to around $10^{-12}$ before $10^5$ yr 
and then quickly increases further.   
As $P$ further increases from 0.01 to 0.03 and 0.05, we can see that the abundance of all COMs and methoxy increase further,  
although generally not as dramatically.

 Now, let us consider the time dependence. Overall, we can see that the abundances of all gas phase species other than formaldehyde
in Fig.~\ref{fig5} decrease at a time near $10^5$ yr and then some species, such as methanol, can recover. 
The decrease in the abundance of glycolaldehyde at this time
can be explained by the depletion of atomic C in gas phase,  as explained in the previous subsection.
The  slight decrease in the abundance of gas phase methanol 
can be explained by the depletion of CO  and gas phase atomic C, which is the major destroyer of methanol in the gas phase.   
Afterwards, as gas phase species such as O, which can react with atomic H on grain surfaces, 
are further depleted from the gas phase, it becomes more likely that atomic H can hydrogenate CO to form methanol instead of reacting
with species such as O. 
Moreover, as gas phase C is depleted, the major destruction process of methanol no longer exists.
Thus, the abundance of methanol both on the topmost layer of the ice mantle and in the gas phase increases again.
Since methanol is the major precursor of methoxy both on the grain surface and in the gas phase, 
the abundance of methoxy as a function of time generally follows that of
methanol. The decrease of gas phase methyl formate and dimethyl ether abundances are the outcomes of the decrease of
the gas phase methoxy abundance and the depletion of atomic C in the gas phase. We can similarly  explain the abundance of gas-phase
ethanol and acetaldehyde as a function of time.
Table~\ref{table4} lists the dominant synthetic routes for gaseous COMs - gas-phase or grain-surface  - for Models M02, M12, M22 and M32 
at a time of $5 \times 10^4$ yr.  It can be seen that a small non-zero value for the probability of reactive desorption is all that is 
needed for the efficient production of the COMs, even though acetaldehyde is formed by a sequence of gas phase reactions following reactive 
desorption, showing that without desorption of suitable precursors, even gas-phase syntheses may not be efficient.   
 Acetaldehyde can also be formed on dust grain surfaces; however, its formation by surface reactions  followed by reactive desorption
only accounts for about 5\% of its total production.

\section{Comparisons with Observations and Previous Models}

It is interesting to see how well our simulation results for COMs and the related radical species agree with observations and the results of previous models. 
The abundances of COMs in our model M02 are too low to to be interesting for comparison with observation, so
they are excluded from our discussion below.  There are virtually no methyl formate molecules formed in model M40  and M431, so
these two models are also excluded from our discussion.
Moreover, no numerical values for abundances of species in the models of ~\citet{Balucani2015} are reported, so
model results in ~\citet{Balucani2015} are only compared with our results briefly.  
We focus on comparisons involving the abundances of species in the sources L1689b and B1-b because only the upper limits of 
the abundances of methoxy, methyl formate and dimethyl ether in L1544 are available~\citep{Vastel2014}.

The comparison with observed values should be made using simulated
abundances at the time of best agreement.
In order to find this time, one approach is to calculate the confidence parameter $k_i$ for each species $i$~\citep{Garrod2007}.  The best agreement
is then obtained when $\sum_{i} k_i$, where the sum is for species detected in the astronomical sources, has its maximum value. 
  For rate equation simulations,
$k_i$ is calculated from the equation
\begin{equation}
k_i= \mathrm{erfc} \left(\frac{|log(R_i)-log(R_i^{'})|}{\sqrt{2}\sigma} \right),
\end{equation}
where $erfc$ is the complementary error function, $R_i$ is the simulated fractional abundance of species $i$, $R_i^{'}$ is the observed abundance
of species $i$ and $\sigma$ is the standard deviation, which is set to be unity here because we assume  
the uncertainty of observed abundances to be one order of magnitude. The value of $k_i$ increases as the ratio $\frac{R_i}{R_i^{'}}$ approaches unity. 
However, for UMMMC simulation results, we cannot directly use the above equation to calculate $k_i$ because the minimum non-zero fractional 
abundance of species we can calculate in our models is $4\times10^{-13}$, so that any fractional abundance  smaller 
than $4\times10^{-13}$ will be 0 by our MC simulation.
The reason why we set the
gas phase fractional abundance resolution to be $4\times10^{-13}$ is clear now, because the smallest observed fractional abundance of the species of interest is that of CH$_3$O, which is $4.7\times 10^{-12}$ as shown in Table~\ref{table5}.  
So, in order to find the best fitting time, we first
find time points, $t_j$, when the number of  species with calculated abundances that differ from their observed values by less than one order of magnitude is a maximum.
If there is only one time point $t_j$, then $t_j$ is the time of best fit. If there is more than one time point, $t_j$,
we then use the above equation to calculate $\sum_m k_m$ at all time points $t_j$, where the sum is only over those species whose calculated abundances differ from 
their observed values by less than one order of magnitude. The best fitting time is the time point $t_n$ when $\sum_m k_m$ is a maximum.  
Our new method to find the best fitting time essentially does not distinguish between bad values and very bad values. One obvious advantage is that
the chosen best fitting time will have the maximum number of species with calculated abundances differing from observed values by less than one
order of magnitude. The species in Table~\ref{table5} are used for analysis. 
We focus on the time range $5\times 10^4$-$2\times 10^5$ yr to find the best agreement time since the so-called early
time when calculated abundances of most species agree well with observations lies around $10^5$ yr for cold cores. 

Table~\ref{table5} shows the best-time simulated abundances of gaseous COMs and methoxy in our models M12, M22, M32, M41, M42, M43, M432 as well as the
models M1 and M10 of \citet{Vasyunin2013a} and model B of \citet{Ruaud2015}, along with observed values 
in the cold cores L1689b and B1-b.  The molecules are discussed individually. 
\begin{itemize}
\item Methanol:  The observed abundance of methanol in B1-b is reasonably fit by the values of previous models except Model B. 
Because of the inclusion of the destruction of gas-phase methanol by atomic C, it is now difficult to reproduce 
the observed methanol abundance with our models except for models M43 and M432. 
The time of best fits in models M43 and M432 is later than that in other models 
so that gas phase C is more depleted at the time of best fits in these two models.
\item Acetaldehyde:  The abundance of this species in B1-b is fit reasonably by all models except models M43 and M432, i
which overproduce it by 1-1.5 orders of magnitude, while models M22 and M32 fail
to reproduce the acetaldehyde abundance in L1689b to within an order of magnitude.
\item Methoxy:  Detected only in B1-b, methoxy has an observed abundance that is reproduced by all of our models.  
The fractional abundances of methoxy in models M1 and M10 \citep{Vasyunin2013a}
are more than one order of magnitude larger than the observed value while the result in model B ~\citep{Ruaud2015}
can fit the observation reasonably well.  \citet{Vasyunin2013a} do not distinguish between methoxy and hydroxymethyl.

\item Methyl formate:  It is difficult to reproduce the high abundance of methyl formate in L1689b with any models 
in which the reactive desorption efficiency $P$ is  less than 0.1.  This statement includes our models M12, M22, and M32, 
as well as models M1 and B.  For our models in which $P=0.1$ (M41, M42, M43 and M432), 
the observed abundance in L1689b can be reproduced only in model M43 and M432 in which $RP=1$, 
but even here the discrepancy is nearly an order of magnitude.  For model M10, in which $P=0.1$, 
the observed abundance in L1689b cannot be reproduced.
It is easer to understand the lower observed abundance of methyl formate in B1-b.  
Our models M12, M22,  M32 and model B contain abundances in agreement with the observed values in B1-b, 
while  model M1 ($P=0.01$) still fails to reproduce the observed abundance.
All models in Table~\ref{table5} with a higher reactive desorption efficiency of $P=0.1$
 can mimic the observed abundance of HCOOCH$_3$.

\item  Formaldehyde: The abundances of this species calculated with model M10 are more than two orders of magnitude larger than the 
observed values in both B1-b and L1689 while the model B result is more than one order of magnitude
larger than the observed value in B1-b. Our model results fit L1689b observations well, however, Models M43 and M432 overestimate
the formaldehyde abundance in B1-b.

\item Dimethyl ether: Model B ~\citep{Ruaud2015}, which sets the reactive desorption efficiency $P$ to be 0.01,  
can reproduce the abundance of CH$_3$OCH$_3$ in L1689b while
all other models with $P=0.01$ fail. This stems from a very high computed value for the abundance of surface 
CH$_{3}$OCH$_{3}$, which, if correct, might be detectable by infrared absorption.  Otherwise, 
the models that produce reasonable fits to the methyl formate abundance also reproduce observed values for dimethyl ether.

\end{itemize}

Overall, only our models M43 and M432 can reproduce the abundances of all species in L1689, but
these two models fail to reproduce acetaldehyde and formaldehyde abundances in B1-b.

\begin{table}
\caption{Fractional abundances of COMs and methoxy in cold cores and time of best fits}
\label{table5}
\begin{tabular}{lllllllll}
  \hline
  source &  Observation       & time   & HCOOCH$_3$ &  CH$_3$OCH$_3$  & CH$_3$CHO   & CH$_3$O  & CH$_3$OH  & H$_2$CO \\ 
         &  or Model          & (yr)  &            &                 &             &          &   \\ \hline
         & Observation        &        & 7.4(-10)   & 1.3(-10)        & 1.7(-10)    &   -      & -       & 1.3(-9)\\
         & M1\tablenotemark{a}          & 1.3(5) & 4.9(-16)   & 2.4(-17)        & 3.2(-11)    &5.3(-14)  &5.7(-12) & 9.8(-10)\\
         & M10\tablenotemark{a}         & 5.1(5) & 3.3(-12)   & 1.3(-10)        & 6.4(-11)    &8.5(-10)  & 2.3(-8) & 5.4(-8)\\
         & Model B\tablenotemark{b}      & 2.0(5) & 3.6(-13)   & 2.35(-10)       & 1.8(-10)    & 7.55(-12) & 3.1(-10) & 1.05(-8)\\ 
 L1689b  &  M41               &  5.3(4) &  6.8(-12) & 3.9(-11) &  2.3(-11)  & 2.0(-12) &  1.6(-10) &  2.6(-9) \\          
         &  M42               & 6.3(4) &  3.0(-11)  &  1.5(-10)   & 1.8(-11) &  1.0(-11) &  1.9(-10) &  3.2(-9) \\                  
         &  M43               & 5.3(4) &  9.4(-11)  &  4.7(-10)  &  1.9(-11) &  2.7(-11) &  2.5(-10) &  2.8(-9)\\                  
         &  M432              &  7.6(4) & 7.8(-11) &  3.5(-10)  &  2.9(-11)  &  2.5(-11) &  3.7(-10) &  5.5(-9)  \\                  
         &  M12\tablenotemark{c}& 1.9(5)& 0         & 0         &  1.7(-10) &   3.4(-11) &  1.1(-9) &  9.5(-9)  \\                  
         &  M22               & 5.3(4)&  1.2(-11)  & 5.0(-11)  &  2.8(-12) &  3.2(-12)   &  3.6(-11) &  9.2(-10) \\ 
         &  M32               & 5.3(4)  & 1.9(-11) &  9.7(-11) &  1.0(-11)  & 5.6(-12) &  7.1(-11) &  1.3(-9) \\  \hline
         &  Observation       &        & 2.0(-11)   &  2.0(-11)       & 1.0(-11)    & 4.7(-12) & 3.1(-9) & 4.0(-10)\\
         &  M1\tablenotemark{a}        & 1.0(6) & 4.2(-15)   & 1.8(-12)        & 4.7(-12)    & 8.5(-11) & 2.5(-9) & 2.9(-9)\\
         &  M10\tablenotemark{a}        & 2.6(5) & 2.0(-12)   & 3.7(-12)        & 3.7(-11)    & 1.5(-10) & 3.3(-9)& 4.8(-8)\\
         &  Model B\tablenotemark{b}      & 5.0(5) & 1.7(-13)  & 5.5(-12)        & 4.7(-12)    & 1.2(-11) & 2.4(-10)& 8.5(-9) \\
 B1-b    &  M41               &  5.3(4) &  6.8(-12) &  3.9(-11) & 2.3(-11) &  2.0(-12) & 1.6(-10) &  2.6(-9) \\                  
         &  M42               &  6.3(4) &  3.0(-11) &  1.5(-10) & 1.8(-11) & 1.0(-11) &  1.9(-10) &  3.2(-9) \\                  
         &  M43               &  9.1(4) &  2.5(-11) &  9.8(-11) & 1.4(-10) &  8.4(-12) &  7.0(-10) &  2.0(-8) \\                  
         &  M432        &  1.1(5) &  1.6(-11) &  6.8(-11) &  4.8(-10) & 3.7(-11) &  1.6(-9) &  3.4(-8) \\                  
         &  M12              &  5.3(4) &  7.2(-12) &  1.1(-11) &  2.4(-12) &  1.2(-12) &  1.3(-11) &  5.9(-10) \\                  
         &  M22               &  5.3(4) &  1.2(-11) &  5.0(-11) &  2.8(-12) &  3.2(-12) &  3.6(-11) &  9.2(-10)\\   
         &  M32               &  5.3(4) &  1.9(-11) &  9.7(-11) &  1.0(-11) &  5.6(-12) &  7.1(-11) &  1.3(-9) \\   
\hline
\end{tabular}
\tablecomments{ Observational values are taken from ~\citet{Vasyunin2013a}. a(b) denotes $a \times 10^b$.}
\tablenotetext{a}{\citet{Vasyunin2013a}. $P$ is  0.1 and 0.01 in models M10 and M1 respectively.}
\tablenotetext{b}{\citet{Ruaud2015}. $P = 0.01$.}
\tablenotetext{c}{Fractional abundances computed to be 0 should be interpreted as $< 4\times10^{-13}$.}
\end{table}

There were also attempts to detect ethanol in B1-b~\citep{Cernicharo2012}. However, ethanol has not been detected 
with a $3\sigma$ upper limit to its column density. We estimate the upper limit of the abundance of ethanol to be $6.7\times 10^{-12}$.  
At the time of best fit for B1-b,  because ethanol can be efficiently formed on grain surfaces in our models other than M432, 
simulated ethanol abundances in these models are around $10^{-10}$ which is more than one order of magnitude larger than 
the upper limit of ethanol abundance in B1-b. On the other hand,  because the surface synthesis of ethanol almost ceases due to 
the depletion of atomic gaseous C in model M432 at the
time of best fit, our simulated ethanol abundance drops to $4.5\times 10^{-11}$, which is still somewhat larger than 
the upper limit to the abundance in B1-b. Since our simulated ethanol abundances are above the observed
upper limit, some important destruction mechanisms may be missing. One possibility is the destruction reaction with atomic carbon, which, although unstudied to the best of our knowledge, may be  rapid because the analogous reaction involving methanol
is rapid at low temperatures~\citep{Shannon2014}. If we assume the rate coefficient of C + C$_2$H$_5$OH 
to be the same as that of  C + CH$_3$OH at 10 K, i.e., about $10^{-10}$ cm$^{3}$ s$^{-1}$, then we can make an order of magnitude estimation of the
significance of the reaction C + C$_2$H$_5$OH. The 
ion-ethanol destruction routes are the major destruction mechanism in our reaction network.  Because ethanol is polar,
we estimate the ion-ethanol rate coefficient to be 10$^{-8}$ cm$^{3}$ s$^{-1}$. The total fractional abundance of 
destructive ions is about $10^{-8}$. So if the fractional abundance of C is greater  than $10^{-6}$ the C + C$_2$H$_5$OH reaction is the
dominant destruction mechanism. From Fig.~\ref{fig3} we can see that the  C + C$_2$H$_5$OH reaction may be the major
destruction mechanism before $10^5$ yr because gas phase C is sufficiently abundant  before this time. However, after $10^5$ yr, the ion-ethanol
reactions remain the major destruction routes of ethanol because gas phase C is severely depleted. 
Our crude estimation can be confirmed by MC simulations.
We included the reaction C + C$_2$H$_5$OH $\rightarrow$ HCO + C$_2$H$_5$
in the reaction network once again assuming the rate coefficient of C + C$_2$H$_5$OH to be the same as that of C + CH$_3$OH at 10 K. We then
ran models M43 and M432 again. The times of best fit are not altered by the new reaction. We found that at these times
($9.1\times 10^{4}$ and $1.1\times 10^{5}$ yr for models M43 and M432 respectively), 
the abundance of ethanol drops to about 1.2$\times 10^{-12}$ in model M43, which is below the observed upper limit
while the abundance of ethanol in model M432 drops only slightly to  1.7$\times 10^{-11}$,  which is still above the observed upper limit
 in B1-b.   Nevertheless, the inclusion of the reaction between C and C$_{2}$H$_{5}$OH improves the agreement with observation.

  Recently \citet{Vastel2014} studied COMs in L1544.  The fractional abundances of methanol and acetaldehyde were found to be 
$\approx 6 \times 10^{-9}$ and $1\times 10^{-10}$ respectively.  
In addition, upper bounds were estimated for methoxy ($1.5\times 10^{-10}$), methyl formate ($1.5\times 10^{-9}$), and dimethyl ether ($2\times 10^{-10}$).  
The standard model of ~\citet{Balucani2015},  which sets $P=0.01$ and includes  new gas phase COM formation reactions,   
can reproduce the observed results in L1544 very well, 
while model B ~\citep{Ruaud2015} can also reproduce observed results in L1544 except that the methanol abundance
is about one order of magnitude lower than the observed value.  At the times of best fit for
B1-b, our models M432 and M43 can reproduce the methanol and acetaldehyde abundances in L1544. 
Moreover, the abundances of methoxy,  methyl formate and dimethyl ether in models M432 and M43  
lie below the upper limits of abundances of these species in L1544. 

It was also recently found that there is a strong correlation between the abundances of 
methyl formate and dimethyl ether~\citep{Jaber2014}.  Specifically,  the abundance of methyl formate is almost 
linearly dependent on the abundance of dimethyl ether in various astronomical sources ranging over five orders of magnitude.  It is argued
that they may have the same precursor or one species may be the precursor of the other one. 
The standard model of ~\citet{Balucani2015} supports the second hypothesis if these two COMs are  formed solely in the gas phase.  
Our models suggest that if they are formed on grain surfaces, then they have the same precursor.
The radical CH$_2$ is a precursor of dimethyl ether while CH is the precursor of methyl formate in our models. 
Dimethyl ether is formed by the gradual hydrogenation of carbon atoms on top of
CH$_3$O while methyl formate is formed by hydrogenation of carbon atoms to CH on top of
CH$_3$O  followed by the reaction between CH and O. The radical CH
is a precursor of CH$_2$. Moreover, the abundances of radicals CH and CH$_2$ are both dependent on gas phase C, so the strong correlation
between the abundances of methyl formate and dimethyl ether can be explained via the primal precursor C. 
 It is also found ~\citep{Jaber2014}
that the correlation between the abundances of 
methyl formate and acetaldehyde is weak, which can be explained by the fact that the  gas-phase synthesis of methyl formate  
requires different precursors from the gas phase synthesis of  acetaldehyde.  Moreover, the formation mechanisms  of acetaldehyde
and methyl formate are not dependent on each other in the gas phase.  

\section{Conclusions}
We performed UMMMC simulations with an efficient reactive desorption 
mechanism under physical conditions pertaining to cold cores, where COMs were 
recently found.  The observed abundances of gas phase COMs and the methoxy radical in the cold 
cores L1689b and  L1544 can be reproduced by our simulations at temperatures as low as 10 K
while most observed gas phase COM abundances in B1-b can also be reproduced without overestimating the
methoxy abundance.  
Moreover, our models can also explain the strong correlation between the abundances of HCOOCH$_3$ and CH$_3$OCH$_3$
and the weak correlation between the abundances of HCOOCH$_3$ and CH$_3$CHO. 

The problem of COM formation in cold cores is linked to the problem of CO$_2$ formation on cold dust grains in our models.   COMs can be formed by the same type of chain reactions used to produce CO$_{2}$ under cold conditions where diffusion of large species does not occur.
The CO$_2$ formation on cold dust grains by chain reactions was
discussed in the Introduction.  Moreover, the formation of surface methoxy on the topmost
layer of an ice mantle also relies on this non-diffusive process. The success of our models
further shows that 
non-diffusive chemical reactions should be included in surface reaction networks for better astrochemical modeling. 
Such processes include our chain reaction mechanism as well as the Eley-Rideal and complex mechanisms advocated by \citet{Ruaud2015}. 
Moreover, reactive desorption plays a key role in the non-thermal desorption of large molecules formed on dust grains. 
Successful experiments on this process are adding to our understanding \citep{Dulieu2013,Minissale2014}.
Experimental results from~\citet{Minissale2014} show that reactive desorption efficiency 
is dependent on surface coverage and varies significantly depending on each specific reaction. 
On the other hand, the reactive desorption efficiency of chain reactions
has never been reported. Chain reactions are different from single reactions in that they are combination of two reactions,
which might enhance reactive desorption efficiency because the energy released by two reactions should be more than 
that by one single reaction. This effect could be important if the reactive desorption efficiency is low for a single
reaction.  
More experimental work on reactive desorption should be undertaken to help explain COM formation in cold astronomical sources.

\begin{acknowledgements}
 We thank the anonymous referee for constructive criticism.  Q. Chang is a research fellow of the One-Hundred-Talent project of the Chinese Academy of Sciences.
E. Herbst acknowledges the support of the National Science Foundation for his astrochemistry program, 
and support from the NASA Exobiology and Evolutionary Biology program through a subcontract from Rensselaer Polytechnic Institute.
\end{acknowledgements}

\end{document}